\documentclass[12pt]{article}
\titlepage

\title{The Lie Algebra of Local Killing Fields}
\author{  Richard Atkins}
\date{}
\newtheorem{fact}{Fact}

\newtheorem{lemma}[fact]{Lemma}
\newtheorem{proposition}[fact]{Proposition}
\newtheorem{theorem}[fact]{Theorem}
\newtheorem{corollary}[fact]{Corollary}
\begin{document}
\maketitle
\begin{abstract} 
We present an algebraic procedure that finds the Lie algebra of the local 
Killing fields of a smooth metric. In particular, we determine the number of
independent local Killing fields about a given point on the manifold.
Spaces of constant curvature, locally symmetric spaces and surfaces are also discussed.
\end{abstract}

\newpage

\section{Introduction}

Killing fields describe the infinitesimal isometries of a metric and as such play a significant
role in differential geometry and general relativity. In this paper we present an algebraic
method that finds the Lie algebra of the local Killing fields of a smooth 
metric $g$. 
In particular, we determine the number of independent local Killing fields of $g$ about 
any given point. In the section following, we identify the local Killing fields of a metric
with local parallel sections of an associated vector bundle $W$, 
endowed with a connection $\nabla$. An examination of the form of the curvature of $\nabla$
leads to a characterization of spaces of constant curvature by means of  a 
system of linear equations. 
In Section 3 we investigate the Lie algebra structure of Killing fields. It is shown that
if the Riemann curvature vanishes at some point on the manifold then the Lie algebra of
Killing fields is isomorphic to a subalgebra of the Lie algebra of the group of isometries
of Euclidean space. Section 4 includes an overview of the procedure developed in \cite{aa}. 
Therein the bundle generated by 
the local parallel sections of $W$ is found by calculating a derived flag of subsets 
of $W$. The number of independent Killing fields of $g$ about a point $x\in M$
is then equal to the dimension of the fibre $\widetilde{W}_{x}$ over $x$ of the terminal subset 
of the derived flag. Associated to $\widetilde{W}_{x}$ is a Lie algebra canonically 
isomorphic to the Lie algebra ${\cal K}_{x}$ of local Killing fields about $x$. 
The method is illustrated  by providing a short proof of a classical theorem that gives
a necessary condition for a space to be locally symmetric, expressed by the vanishing of a 
set of quadratic homogeneous polynomials in the curvature. 
Section 5 considers the derived flag for Riemannian surfaces. We obtain a classification
of the Riemannian metrics corresponding to the various possible 
kinds of Lie algebra ${\cal K}_{x}$.

\section{Killing Fields and Constant Curvature}

We associate to Killing fields parallel sections of a 
suitable vector bundle in the manner put forward by Kostant (cf. \cite{ff}). 
The utility of such a framework is two-fold: first, it
permits us to apply algebraic techniques adapted to finding the  
subbundle generated by local parallel sections. Second, it enables a purely algebraic 
description of the Lie bracket of two Killing fields, avoiding the explicit appearance of 
derivatives. 

Let $g$ be a metric on a differentiable manifold $M$ of dimension $n$; $g$ is assumed to be
pseudo-Riemannian of signature $(p,q)$ unless otherwise stated.
$K$ is a Killing field of $g$ if and only if
\begin{eqnarray} \label{killing1}
 K_{a;b} + K_{b;a} = 0 
\end{eqnarray}
where the semi-colon indicates covariant differentiation with respect to the Levi-Civita
connection of $g$. It is straightforward to verify that
\begin{eqnarray} \label{killing2}
 K_{a;bc} = R_{abc}^{\hspace{.2in}d}K_{d}
\end{eqnarray}
for Killing fields $K$, where ${R_{abc}}^{d}$ is the Riemann curvature tensor 
of $g$, defined according to 
\[ A_{c:ba}-A_{c:ab} = {R_{abc}}^{d}A_{d} \]
The summation convention shall be used throughout.

Let $W$ be the Whitney sum $W:= T^{*}M \oplus \Lambda^{2}T^{*}M$. A local section of $W$ 
has the form $X=K+L$, where $K=K_{a}dx^{a}$ is a local section of $T^{*}M$ 
and $L=L_{ab}dx^{a}\wedge dx^{b}$ is
a local section of $\Lambda^{2}T^{*}M$. Define a connection $\nabla$ on $W$ by 
\begin{eqnarray} \label{parallel}
\nabla_{i} X = (K_{a;i}-L_{ai})dx^{a} +
(L_{ab;i} - {R_{abi}}^{c}K_{c})dx^{a}\wedge dx^{b} 
\end{eqnarray} 
For an open subset $U\subseteq M$, let ${\cal K}_{U}$ denote the local Killing fields 
$K:U\longrightarrow T^{*}M$ and let ${\cal P}_{U}$ denote the local parallel sections
$X:U\longrightarrow W$; the subscript $U$ shall be omitted when $U=M$.
Define the map $\phi_{U}:{\cal K}_{U} \longrightarrow {\cal P}_{U}$ by 
\begin{eqnarray}
\phi_{U}(K_{a}):= K_{a}+K_{a;b}
\end{eqnarray}
It is clear that the image of $\phi_{U}$ does, in fact, lie in  ${\cal P}_{U}$.
The inverse $\psi_{U}:{\cal P}_{U}\longrightarrow {\cal K}_{U}$ of $\phi_{U}$ 
is the projection of $W$ onto $T^{*}M$: $\psi_{U}(K_{a}+L_{ab}):=K_{a}$.
This establishes a vector space isomorphism
\begin{eqnarray}
{\cal K}_{U} \hspace{0.2in} \leftrightarrow  \hspace{0.2in} {\cal P}_{U}
\end{eqnarray}

Consider a vector space $V$ with a non-degenerate, symmetric  bilinear form $h$. 
Let $B=B_{abcd}$ be a covariant $4$-tensor on $V$ satisfying  
the following relations common to a Riemann curvature tensor:
\begin{eqnarray} \label{sym}
B_{abcd}=B_{cdab}=-B_{bacd}=-B_{abdc} 
\end{eqnarray}
and let $T=T_{ab\cdots}$ be an $n$-tensor on $V$ with $n\geq 2$. The derivation $B\star T$ 
is the $(n+2)$-tensor defined by
\begin{eqnarray}
{B\star T}_{abcd\cdots} := B_{sbcd}{T^{s}}_{a\cdots}+B_{ascd}{T^{s}}_{b\cdots}+
           B_{absd}{T^{s}}_{c\cdots}+B_{abcs}{T^{s}}_{d\cdots}
\end{eqnarray}
Indices are raised by $h$.

\begin{lemma} \label{2diml}
If $V$ is 2-dimensional then $B\star L=0$ for all $L\in \Lambda^{2}V^{*}$.
\end{lemma}
{\bf Proof:}\\
It shall be convenient to work in an orthonormal basis of $V$ in which  
$h=diag(\eta_{1},\eta_{2})$, where $\eta_{i} =\pm 1$. Then ${L^{i}}_{j} = \eta_{i}L_{ij}$. 
Owing to the symmetries (\ref{sym}), there are effectively two cases to consider. 
\vspace{0.1in} \\
(i) $a=b=1$ case: 
\begin{eqnarray}
B\star L_{abcd} = \eta_{2}(B_{21cd}+B_{12cd})L_{21}=0 \nonumber 
\end{eqnarray}
(ii) $a=c=1$, $b=d=2$ case:
\begin{eqnarray}
B\star L_{abcd} = \eta_{2}B_{2212}L_{21}+\eta_{1}B_{1112}L_{12}+\eta_{2}B_{1222}L_{21}+
                  \eta_{1}B_{1211}L_{12} =0 \nonumber
\end{eqnarray}
{\bf q.e.d.}

Applying the Bianchi identities, the curvature $F(i,j):= \nabla_{[i}\nabla_{j]}$ of $\nabla$
takes the form 
\begin{eqnarray} \label{curvature}
 F(i,j)(X) = (R_{ijkl;s}K^{s} +  {R\star L}_{ijkl})dx^{k}\wedge dx^{l}
\end{eqnarray}
where $X=K_{a}dx^{a} +L_{ab}dx^{a}\wedge dx^{b}$ (cf. \cite{cc}).
In the sequel, it shall be convenient to view the curvature $F$ as a map 
$F:W\longrightarrow \Lambda^{2}T^{*}M\otimes W$
given by $w \mapsto F(,)(w)$. $F$ is composed of two pieces: a 
$K$-part and an $L$-part. The $K$-part provides a description of locally symmetric spaces:
$g$ is locally symmetric if and only if $T^{*}M \subseteq ker \hspace{0.03in} F$.
The $L$-part, on the other hand, provides a characterization of metrics of constant sectional 
curvature by means of a system of homogeneous linear equations.

\begin{proposition} \label{thm:sf}
Let $g$ be Riemannian and $n\geq 3$. Then $M$ is a space of constant curvature if and only if 
\[ R\star L =0 \]
for all  $L\in \Lambda^{2}T^{*}M$.
\end{proposition}

Expressed in terms of indices, $M$ has constant curvature (for $n\geq 3$) if and only if 
for all $L\in \Lambda^{2}T^{*}M$,
\begin{eqnarray} \label{sf}
 R_{sjkl}{L^{s}}_{i}+R_{iskl}{L^{s}}_{j}+
           R_{ijsl}{L^{s}}_{k}+R_{ijks}{L^{s}}_{l} =0 
\end{eqnarray}
It is evident from the lemma that the theorem does not hold for $n=2$.\\  
{\bf Proof}:\\
$\Longrightarrow$ If $g$ has constant curvature then 
$R_{ijkl} = \kappa_{0}(\delta_{il}\delta_{jk}
-\delta_{ik}\delta_{jl})$ with respect to an orthonormal frame, where $\kappa_{0}$ is a
constant. Substitution of this expression into the left hand side of (\ref{sf})
gives zero for all skew-symmetric $L=L_{ab}$. \\ 
$\Longleftarrow$ Suppose that (\ref{sf}) holds for all $L\in \Lambda^{2}T^{*}M$.
We shall work in an orthonormal frame 
$X_{1},...,X_{n}$ for $g$; this will allow us to deal with lowered indices throughout: 
${L^{i}}_{j}=L_{ij}$. 
Let $i=k,j$ and $l$ be three distinct indices in (\ref{sf}). This gives
\begin{eqnarray} \label{i=a}
 R_{sjil}L_{si}+R_{isil}L_{sj}+ R_{ijsl}L_{si}+R_{ijis}L_{sl} =0 
\end{eqnarray}
Put $L_{rs} := \delta_{rl}\delta_{sj}-\delta_{rj}\delta_{sl}$ into (\ref{i=a}) to obtain
$ R_{ijij} = R_{ilil} $.
It follows that for any two pairs of distinct indices $i\neq j$ and $a\neq b$,
$R_{ijij} = R_{abab}$. Thus 
\begin{eqnarray} \label{2indices}
R_{ijij} = \kappa(x) \hspace{0.5in} \mbox{for $i\neq j$}
\end{eqnarray}
where $\kappa$ is some function on $M$.

Next, let $i=k$ and $j=l$ be two distinct indices in (\ref{sf}). This gives:
\begin{eqnarray} \label{3indices}
 R_{ijis}L_{sj} + R_{ijsj}L_{si} = 0  
 \end{eqnarray}
Let $m$ be any index distinct from $i$ and $j$ and put 
$L_{rs} := \delta_{rm}\delta_{sj}-\delta_{rj}\delta_{sm}$ into (\ref{3indices}). We obtain
\begin{eqnarray} \label{i=a,j=b}
R_{ijim} = 0 \hspace{0.5in} \mbox{for $i,j$ and $m$ distinct}
\end{eqnarray}

Consider a pair $Y_{1},Y_{2}$ of orthonormal vectors in $T_{x}M$. If $X_{1},X_{2}$ span
the same plane as $Y_{1},Y_{2}$ at $x$ then $R(Y_{1},Y_{2},Y_{1},Y_{2})=\kappa(x)$, 
by (\ref{2indices}). If
$Y_{1},Y_{2}$ span a plane orthogonal to $X_{1},X_{2}$  then we may as well suppose 
$X_{3}=Y_{1}$ and $X_{4}=Y_{2}$, whence $R(Y_{1},Y_{2},Y_{1},Y_{2})=\kappa(x)$, from 
(\ref{2indices}) again. The last possibility is that $Y_{1},Y_{2}$ and $X_{1},X_{2}$ span planes
that intersect through a line, which for the purpose of calculating sectional curvature
we may take to be generated by $X_{1}=Y_{1}$, by means of appropriate rotations of the
pairs $X_{1},X_{2}$ and $Y_{1},Y_{2}$ within the respective planes they span. We may suppose,
furthermore, that $X_{3}$ is the normalized component of $Y_{2}$ orthogonal to $X_{2}$;
thus $Y_{2}=aX_{2}+bX_{3}$, where $a^{2}+b^{2}=1$. From (\ref{2indices}) and (\ref{i=a,j=b})
this gives
\begin{eqnarray}
R(Y_{1},Y_{2},Y_{1},Y_{2}) & = & R(X_{1},aX_{2}+bX_{3},X_{1},aX_{2}+bX_{3}) \nonumber \\
  & = & a^{2}R_{1212}+b^{2}R_{1313} \nonumber \\
  & = & \kappa(x) \nonumber
\end{eqnarray}
Therefore $g$ has constant curvature at each point $x\in M$. By Schur's Theorem, $g$ has 
constant curvature. \\
{\bf q.e.d.}

\section{The Lie algebra Structure of ${\cal K}_{U}$}

Let $V$ be an $n$-dimensional
vector space equipped with a non-degenerate, symmetric bilinear form $h$,
of signature $(p,q)$, and let $B=B_{abcd}$ be a covariant 4-tensor on $V$ 
satisfying the usual algebraic relations of a Riemann curvature:
\begin{eqnarray}
& B_{abcd} &  =   -B_{bacd}  \label{b1} \\
& B_{abcd} &  =   -B_{abdc}  \label{b2} \\
& B_{abcd} & \hspace{-0.1in} +   B_{acdb}  +   B_{adbc}  =  0, \label{b3}
 \hspace{0.1in} \mbox{and afortiori} \\
& B_{abcd} & = B_{cdab} \nonumber
\end{eqnarray} 
By virtue of (\ref{b1}) and (\ref{b2}) we may define a skew-symmetric, bilinear
bracket operation on 
$V^{*}\oplus \Lambda^{2}V^{*}$ by 
\begin{eqnarray} \label{vbracket}
[K_{a}+L_{ab},{K'}_{a}+{L'}_{ab}] := {L'}_{ab}{K}^{b}-{L}_{ab}{K'}^{b}+
              {{L'}_{a}}^{c}{L}_{cb}- {L_{a}}^{c}{L'}_{cb}+ B_{abcd}K^{c}{K'}^{d} 
\end{eqnarray}
where indices are raised and lowered with $h$. 
If a subspace ${\cal W}$ of $V^{*}\oplus \Lambda^{2}V^{*}$ is closed with respect to
the bracket and satisfies the Jacobi identity then we denote the 
associated Lie algebra by ${\cal A}({\cal W},B,h)$.

\begin{lemma} \label{jacobi}
Let ${\cal W}$ be a subspace of $V^{*}\oplus \Lambda^{2}V^{*}$, closed with respect to
the bracket operation.  The Jacobi identity holds on ${\cal W}$  if and only if for all
$X=K+L, X'=K'+L'$ and $X''=K''+L''$ in ${\cal W}$, where $K,K',K''\in V^{*}$ and 
$L,L',L'' \in \Lambda^{2}V^{*}$, 
\begin{eqnarray} 
B\star L_{abcd}{K'}^{c}{K''}^{d}+
B\star L'_{abcd}{K''}^{c}{K}^{d}+
B\star L''_{abcd}{K}^{c}{K'}^{d} =0
\end{eqnarray} 
\end{lemma}
{\bf Proof:} \\
Let $K,K',K'' \in V^{*}$ and $L,L',L'' \in \Lambda^{2}V^{*}$. 
There are four cases to consider. 
\vspace{0.1in} \\
(i) $K-K'-K''$ case. We have 
\[ [K,[K',K'']]=[K,B_{abcd}{K'}^{c}{K''}^{d}]=B_{abcd}K^{b}{K'}^{c}{K''}^{d} \]
Therefore,
\begin{eqnarray} 
& & [K,[K',K'']] + [K',[K'',K]]+[K'',[K,K']] \nonumber \\
& = & (B_{abcd}+B_{acdb}+B_{adbc}) K^{b}{K'}^{c}{K''}^{d} \nonumber \\
& = & 0       \label{jac1}       
\end{eqnarray}
by equation (\ref{b3}). \vspace{0.1in} \\
(ii) $K-K'-L$ case. First,
\[ [K,[K',L]] = [K,L_{ab}{K'}^{b}] = B_{abcd}K^{c}{L^{d}}_{s}{K'}^{s} \]
Also,
\begin{eqnarray}
[L,[K,K']] & = &  [L,B_{abcd}K^{c}{K'}^{d}] \nonumber \\
           & = &  B_{ascd}K^{c}{K'}^{d}{L^{s}}_{b}-{L_{a}}^{s}B_{sbcd}K^{c}{K'}^{d} \nonumber
\end{eqnarray}
Combining these with (\ref{b2}) and the fact that $L=L_{ab}$ is skew-symmetric, we obtain
\begin{eqnarray} \label{jac2}
[K,[K',L]]+[K',[L,K]]+[L,[K,K']]= B\star L_{abcd} K^{c}{K'}^{d}
\end{eqnarray} 
(iii) $K-L-L'$ case. Observe that
\[ [K,[L,L']] = [K,L'L-LL'] = L'LK-LL'K \]
and 
\[ [L,[L',K]]= -[L,L'K]= LL'K \]
Using these equations gives
\begin{eqnarray} \label{jac3}
[K,[L,L']]+[L,[L',K]]+[L',[K,L]]= 0
\end{eqnarray}
(iv) $L-L'-L''$ case. It is elementary to verify that
\begin{eqnarray} \label{jac4}
[L,[L',L'']]+[L',[L'',L]]+[L'',[L,L']] = 0
\end{eqnarray}
After applying (\ref{jac1})-(\ref{jac4}), 
\[ [X,[X',X'']]+[X',[X'',X]]+[X'',[X,X']] \]
simplifies to
\[ B\star L_{abcd}{K'}^{c}{K''}^{d}+
B\star L'_{abcd}{K''}^{c}{K}^{d}+
B\star L''_{abcd}{K}^{c}{K'}^{d}  \]
{\bf q.e.d.} 

\begin{proposition} 
(i) If $V$ is 2-dimensional then any subspace ${\cal W}$ of 
$V^{*}\oplus \Lambda^{2}V^{*}$, closed with respect to the bracket, defines a Lie algebra
${\cal A}({\cal W}, B, h)$. \\
(ii) If $V$ is arbitrary and $B=0$ then $V^{*}\oplus \Lambda^{2}V^{*}$ defines a Lie algebra 
${\cal A}(V^{*}\oplus \Lambda^{2}V^{*},B=0,h)$. 
\end{proposition}
{\bf Proof:} \\
(i) follows from Lemmas \ref{2diml} and \ref{jacobi}. (ii) is immediate.\\
{\bf q.e.d.}

For local Killing fields $K,K'\in {\cal K}_{U}$, the Lie bracket $K'':=[K,K']$ is
\begin{eqnarray}
{K''}_{a}   =  {L'}_{ab}K^{b} -L_{ab}{K'}^{b}
\end{eqnarray}
where we have written $L_{ab}:=K_{a;b}$ and ${L'}_{ab}:={K'}_{a;b}$.
Furthermore, ${L''}_{ab} :=  {K''}_{a;b}$ is given by
\begin{eqnarray}
{L''}_{ab}& = &  {{L'}_{a}}^{c}L_{cb}-{L_{a}}^{c}{L'}_{cb}
                     +R_{abcd}K^{c}{K'}^{d}
\end{eqnarray}
Defining a bracket on ${\cal P}_{U}$ by 
\begin{eqnarray} \label{pbracket}
[K+L,K'+L'] := {L'}_{ab}{K}^{b}-{L}_{ab}{K'}^{b}+
              {{L'}_{a}}^{c}{L}_{cb}- {L_{a}}^{c}{L'}_{cb}+ R_{abcd}K^{c}{K'}^{d} 
\end{eqnarray}
gives an isomorphism $\phi_{U}:{\cal K}_{U} \longrightarrow {\cal P}_{U}$ of Lie algebras:
\begin{eqnarray} \label{lieiso}
\phi_{U}([K,K']) = [\phi_{U}(K),\phi_{U}(K')] 
\end{eqnarray}

For $x \in U$, define the subspace $W_{U,x}$ of $W_{x}$ by
\begin{eqnarray}
W_{U,x}:= \{w\in W_{x}:w=X(x), \hspace{0.1in} \mbox{for some} \hspace{0.1in} 
               X\in{\cal P}_{U} \}
\end{eqnarray}
Since parallel sections of a vector bundle are determined by their value at a single point,
${\cal P}_{U}$ and  $W_{U,x}$ are isomorphic as vector spaces via the restriction
map $r_{x}: {\cal P}_{U} \longrightarrow W_{U,x}$, given by $r_{x}(X):=X(x)$.
Comparing (\ref{vbracket}) and (\ref{pbracket}), we have, in fact, 
a Lie algebra isomorphism $r_{x}:{\cal P}_{U} \longrightarrow {\cal A}(W_{U,x},R_{x},g_{x})$.
Composing this with $\phi_{U}$ characterizes the Lie algebra of ${\cal K}_{U}$.

\begin{lemma} \label{iso}
\begin{eqnarray} \label{isomorphism}
r_{x} \circ \phi_{U}: {\cal K}_{U} \longrightarrow {\cal A}(W_{U,x},R_{x},g_{x})
\end{eqnarray}
is an isomorphism of Lie algebras.
\end{lemma}

In order to find the Lie algebra of the local Killing fields of $g$ about the point $x$
it remains to calculate $W_{U,x}$ for a sufficiently small neighbourhood $U$ of $x$.
This is accomplished in the following section.

\begin{lemma} \label{sd}
${\cal A}(V^{*}\oplus \Lambda^{2}V^{*},B=0,h)$ is isomorphic to the Lie algebra of
the semidirect product $\Re^{n}\times_{sd} SO(p,q)$.
\end{lemma}
{\bf Proof:} \\
Let $M=U=\Re^{n}$. Identify $V$ and $T_{x}M$, for some fixed
$x\in M$, by a vector space isomorphism $\theta: T_{x}M \longrightarrow V$. 
Set $g_{x} := \theta^{*}(h)$, the pull-back of $h$. $g_{x}$ extends naturally 
to a flat metric $g$ of signature $(p,q)$ defined on all of $M$. 
The Lie algebra of Killing fields ${\cal K}$ is then isomorphic to the Lie algebra of 
$\Re^{n}\times_{sd} SO(p,q)$. Furthermore,
$W_{M,x}=T^{*}_{x}M\oplus\Lambda^{2}T^{*}_{x}M$ and as Lie algebras,
\[ 
 {\cal A}(V^{*}\oplus \Lambda^{2}V^{*},B=0,h) \cong 
{\cal A}(T^{*}_{x}M\oplus\Lambda^{2}T^{*}_{x}M,R_{x}=0,g_{x}) \cong  {\cal K} \nonumber
\]
where the second isomorphism follows from Lemma \ref{iso}.\\
{\bf q.e.d.}

This leads to the following global result.
\begin{proposition} If the Riemann curvature tensor 
vanishes at some point $x \in M$ then the 
Lie algebra of Killing fields is isomorphic to a subalgebra of the Lie algebra of
$\Re^{n}\times_{sd} SO(p,q)$.
\end{proposition}
{\bf Proof:}\\
Suppose $R_{x}=0$. For $U=M$, Lemma \ref{iso} 
gives the isomorphism  ${\cal K}\cong {\cal A}(W_{M,x},R_{x}=0,g_{x})$.  
This is a  subalgebra of ${\cal A}(W_{x},R_{x}=0,g_{x})$, which by Lemma \ref{sd}
is isomorphic to the Lie algebra of $\Re^{n}\times_{sd} SO(p,q)$. \\
{\bf q.e.d.}

\section{Parallel Fields and Locally Symmetric Spaces}

We begin by briefly reviewing the method from \cite{aa}.
This describes an algebraic procedure for determining
the number of independent local parallel sections of a smooth vector bundle
$\pi: W\rightarrow M$ with a connection $\nabla$. Since the existence theory is
based upon the Frobenius Theorem, smooth data are required.

Let $W'$ be a subset of $W$ satisfying the following two properties: \\
P1: the fibre of $W'$ over each $x\in M$ is a linear subspace of the fibre of
    $W$ over $x$, and \\
P2: $W'$ is {\it level} in the sense that each element $ w\in W'$ is contained in the image 
    of a     local smooth section of $W'$, defined in some neighbourhood of $\pi(w)$ in $M$.

Let $X$ be a local section of $W'$. The covariant derivative of $X$ is a local section
of $W\otimes T^{*}M$. Define $\widetilde{\alpha}$ by 
\[ \widetilde{\alpha}(X) : = \phi \circ \nabla(X) \]
where $\phi:W\otimes T^{*}M \rightarrow (W/W')\otimes T^{*}M$ denotes the natural projection
taken fibrewise. If $f$ is any differentiable function with the same domain as $X$ then
\[ \widetilde{\alpha}(fX) = f\widetilde{\alpha}(X) \]
This means that $\widetilde{\alpha}$ defines a map 
\[ \alpha_{W'} : W' \rightarrow (W/W')\otimes T^{*}M \]
which is linear on each fibre of $W'$.

The kernel of $\alpha_{W'}$ is a subset of $W'$, which satisfies property P1 
but not necessarily property P2. In order to carry out the above constructions
to $ker \hspace{0.03in} \alpha_{W'}$, as we did to $W'$, the non-level points in  
$ker \hspace{0.03in} \alpha_{W'}$ must be removed. 
To this end we define a leveling map ${\cal S}$ as follows.
For any subset $V$ of $W$ satisfying P1 let ${\cal S}(V)$ be the subset of
$V$ consisting of all elements $v$ for which there exists a smooth local section
$s:U\subseteq M \rightarrow V\subseteq W$ such that $v = s(\pi(v))$. Then 
${\cal S}(V)$ satisfies both P1 and P2.

We may now describe the construction of the maximal flat subset $\widetilde{W}$,  of $W$. 
Let
\[ \begin{array}{lll}
   V^{(0)} & := & \{ w \in W \hspace{.03in} | \hspace{.03in} F(,)(w) = 0 \} \\
   W^{(i)}   & := & {\cal S}(V^{(i)}) \\
   V^{(i+1)} & := & ker \hspace{.03in} \alpha_{W^{(i)}} 
  \end{array} \]
where, as before, $F:TM\otimes TM\otimes W\rightarrow W$
is the curvature tensor of $\nabla$. This gives a sequence
\[ W \supseteq W^{(0)} \supseteq W^{(1)} \supseteq \cdots \supseteq W^{(k)} 
\supseteq \cdots \]
of subsets of $W$. 
For some $k \in { N}$, $W^{(l)}= W^{(k)}$ for all $l \geq k$. 
Define $\widetilde{W} = W^{(k)}$, with projection $\tilde{\pi}:\widetilde{W} \rightarrow M$. 

We say that the connection $\nabla$ is {\it regular at $x\in M$} if there exists 
a neighbourhood $U$ of $x$ such that $\tilde{\pi}^{-1}(U) \subseteq \widetilde{W}$
is a vector bundle over $U$. 
$\widetilde{W}_{x}$ shall denote the fibre of $\widetilde{W}$ over $x\in M$. 

\begin{lemma} \label{lemma:flag}
Let $\nabla$ be a connection on the smooth vector bundle $\pi:W\rightarrow M$. \\
(i) If $X:U \subseteq M \rightarrow W$ is a local parallel section then the image of $X$ lies in
$\widetilde{W}$. \\
(ii) Suppose that $\nabla$ is regular at $x\in M$. Then for every $w \in \widetilde{W}_{x}$
there exists a local parallel section 
$X:U\subseteq M \rightarrow \widetilde{W}$ with $X(x) = w$.
\end{lemma}

We may now describe the Lie algebra of local Killing fields about a point $x$. 

\begin{theorem} \label{thm:main}
Let $g$ be a smooth metric on a manifold $M$ with associated connection $\nabla$ on 
$W= T^{*}M \oplus \Lambda^{2}T^{*}M$, which is assumed to be regular at $x\in M$. Then
$g$ has  $dim \hspace{0.03in} \widetilde{W}_{x}$ independent local 
Killing fields in a sufficiently small 
neighbourhood $U$ of $x\in M$. Moreover, the Lie algebra of Killing fields 
on $U$ is canonically isomorphic to  the Lie algebra ${\cal A}(\widetilde{W}_{x},R_{x},g_{x})$.
\end{theorem}
{\bf Proof:} \\
By Lemma \ref{lemma:flag} there exists a sufficiently small
open neighbourhood $U$ of $x$ such that
$W_{U,x} = \widetilde{W}_{x}$. The theorem now follows from Lemma \ref{iso}. \\
{\bf q.e.d.}

As an illustration, we provide a short algebraic proof of the classical theorem 
that locally symmetric spaces satisfy
\begin{eqnarray} \label{rr=0}
R\star R=0 
\end{eqnarray}
(cf. \cite{ee}, pg. 197, 2nd and 3rd equations).

\begin{lemma} \label{lemma:cotangent}
If $M$ is locally symmetric then $T^{*}M\subseteq \widetilde{W}$.
\end{lemma}
{\bf Proof:}\\
Suppose $M$ is locally symmetric and let $x\in M$.  Then there is an open neighbourhood
$U$ of $x$ such that the space of Killing fields on $U$,  whose covariant derivative vanishes 
at $x$, has dimension $n$. By the isomorphism given
in Lemma \ref{iso}, $T^{*}_{x}M \subseteq W_{U,x}$.  From Lemma \ref{lemma:flag} (i) we have 
$W_{U,x} \subseteq \widetilde{W}_{x} $ and so $T^{*}_{x}M \subseteq  \widetilde{W}_{x} $.\\
{\bf q.e.d.} 

Let $T=T_{a\cdots}$ be an $n$-tensor with $n\geq 1$.
Define $R\cdot T$ to be the $(n+2)$-tensor
obtained by contracting the rightmost index of $R$ with the leftmost index of $T$:
\begin{eqnarray}
{R\cdot T}_{abc\cdots} := {R_{abc}}^{s}T_{s\cdots}
\end{eqnarray}
Let ${\sf p}:=T^{*}M$, the cotangent space of $M$ and let 
${\sf p}^{(1)}$ be defined as the set of all elements $v\in T^{*}M$ satisfying
\begin{eqnarray}
R\star R\cdot v =  0 
\end{eqnarray}
Define ${\sf t}$ to be the set of all $L\in \Lambda^{2}T^{*}M$ 
such that
\begin{eqnarray} \label{l=0}
R\star L=0
\end{eqnarray}
We shall assume that ${\sf t}$ has constant rank.  \vspace{0.1in} \\
{\bf Proof of (\ref{rr=0}):}\\
Let us calculate the derived flag of $W$ supposing $M$ to be locally symmetric. 
From (\ref{curvature}), 
$W^{(0)} = V^{(0)} := ker \hspace{0.03in} F = {\sf p}\oplus {\sf t}$. 
Let $X=K+L$ be a local section of $W^{(0)}$, where $K$ and $L$ are local sections of 
${\sf p}$ and ${\sf t}$, respectively. By definition, $X(x) \in V_{x}^{(1)}$ 
if and only if $\nabla_{i} X(x) \in W^{(0)}_{x}$, for all $i$. This is equivalent to 
\begin{center}
$(*):$ \hspace{0.1in} $L_{ab; (i)}-{R_{ab(i)}}^{c}K_{c} \in {\sf t}$ \hspace{0.1in} at $x$. 
\end{center}
Taking the covariant derivative of $R\star L=0$ gives $R\star L_{;(i)}=0$.
Thus $L_{ab;(i)}$ is a local section of ${\sf t}$.
This means that $(*)$ holds if and only if $R\cdot K_{x}$ lies in ${\sf t}$. 
This is the case precisely when $K \in {\sf p}^{(1)}$ at $x$, and so
$V^{(1)} = {\sf p}^{(1)} \oplus {\sf t}$. By Lemma \ref{lemma:cotangent},
we must have $V^{(1)} = T^{*}M \oplus {\sf t}$, whence  it follows that ${\sf p}^{(1)}=T^{*}M$
and $W^{(1)}=V^{(1)}=W^{(0)}$. Therefore $R\star R=0$. \\
{\bf q.e.d.} 

We have also shown that for $M$ locally symmetric,  
$\widetilde{W}=W^{(0)}={\sf p} \oplus {\sf t}$. 
Conversely, if 
$\widetilde{W}={\sf p} \oplus {\sf t}$ then $T^{*}M \subseteq ker \hspace{0.03in} F$,
from which we conclude that $M$ is locally symmetric. The terminal subbundle of the derived 
flag therefore characterizes locally symmetric spaces:
\begin{eqnarray} \label{lss}
\mbox{ $M$ is a locally symmetric space if and only if $\widetilde{W}={\sf p} \oplus {\sf t}$.}
\end{eqnarray}
In particular, the derived flag computes the local canonical decomposition.

As observed above, a Riemannian manifold  with $dim \hspace{0.05in}M \geq 3$
has constant curvature if and only if $R\star L=0$ for all $L\in \Lambda^{2}T^{*}M$
(Proposition \ref{thm:sf}). Furthermore, all manifolds of dimension $n=1$ or $2$ 
satisfy  $R\star L=0$ for $L\in \Lambda^{2}T^{*}M$ (cf. Lemma \ref{2diml}). 
Since spaces of constant curvature are locally symmetric
the curvature $F$ vanishes for such manifolds. It is not difficult to see that the
converse holds (for the 2-dimensional case use the canonical form of the Riemann
curvature: $R_{ijkl} = c(g_{il}g_{jk}-g_{ik}g_{jl})$, where $c$ is the 
Gaussian curvature). Consequently,
\begin{eqnarray} \label{scc}
\mbox{ $M$ is a space of constant curvature if and only if $\widetilde{W}=W$. }
\end{eqnarray}
Employing Theorem \ref{thm:main}, we obtain as a corollary the familiar result that a 
Riemannian manifold possesses the maximal possible number $\frac{1}{2}n(n+1)$ of independent 
local Killing fields if and only if it is a space of constant curvature.

\section{Classification for Riemannian Surfaces}

As a final illustration of Theorem \ref{thm:main}, we shall 
determine which Riemannian surfaces correspond to the various types of
Lie algebra ${\cal K}_{x}$. 
In the process, a necessary condition for a Riemannian surface to possess a 
Killing field is obtained.
First, we recall the situation involving the  maximal number of local Killing fields:\\
{\it Let $M$ be a Riemannian surface. The following are equivalent: \\
\hspace*{0.2in}(i) \hspace*{0.06in} $dim \hspace{0.05in} {\cal K}_{x} =3$ for all $x\in M$. \\
\hspace*{0.2in}(ii) \hspace*{0.01in} $M$ has constant Gaussian curvature $c$. \\
\hspace*{0.2in}(iii) $M$ is locally symmetric.\\
In this case, \\
\hspace*{0.2in}(a)  if $c=0$ then ${\cal K}_{x}$ is isomorphic to the Lie algebra of 
                         $\Re^{2}\times_{sd}SO(2)$; \\
\hspace*{0.2in}(b)  if $c>0$ then ${\cal K}_{x}\cong {\sf sl}_{2}\Re$; \\
\hspace*{0.2in}(c)  if $c<0$ then ${\cal K}_{x}\cong {\sf su}_{2}$}.

The equivalence of (i)-(iii) follows from Lemma \ref {2diml}, 
(\ref{scc}) and the observation below 
(\ref{scc}). Assume that these conditions hold. 
$c=0$ corresponds, locally, to flat Euclidean space, for which ${\cal K}_{x}$ is isomorphic to 
the Lie algebra of the semidirect product of translations and rotations.
Suppose $c\neq 0$ and let $X$ and $Y$ be orthogonal vectors in $T_{x}^{*}M$ with norm
\[ X^{2} = Y^{2} = \frac{1}{|c|} \]
Define $H \in \Lambda^{2}T^{*}_{x}M$ by
\[ [X,Y] := H \]
Then 
\[ [H,X] = -sg(c)Y \hspace{0.5in} \mbox{and} \hspace{0.5in} [H,Y] = sg(c)X \]
where $sg(c)$ denotes the sign of $c$. Appealing to Lemma \ref{iso},
this identifies ${\cal K}_{x}$ with ${\sf sl}_{2}\Re$ 
for $c>0$ and with ${\sf su}_{2}$ for $c<0$ (cf. \cite{dd}, pg. 143).  

Next, we calculate the derived flag for $W$ assuming that the surface is
{\it regular}: $W^{(i)} = V^{(i)}$ has constant rank for all $i$. 
The elements $K\in T^{*}M$ satisfying
${R_{ijkl}}^{;s}K_{s}$ are those for which $c^{,s}K_{s} = 0$. 
Therefore, by Lemma \ref{2diml},
\begin{eqnarray}
W^{(0)} = ker \hspace{0.03in} \partial c \oplus \Lambda^{2}T^{*}M 
\end{eqnarray}
where $\partial c$ denotes the vector field $c^{,a}$.
The case  $\partial c =0$ has been handled above. 
Suppose therefore that $dim \hspace{0.03in} ker \hspace{0.03in} \partial c =1$; that is, 
$\partial c$ is non-vanishing. 
Then $W^{(0)}$ is a rank two fibre bundle over $M$. 
Let $X=K+L$ be a local section of $W^{(0)}$, where $K$ is a local section of 
$ker \hspace{0.03in} \partial c$ and $L$ is a local section of $\Lambda^{2}T^{*}M$.
$X(x) \in W_{x}^{(1)}$ is equivalent to  $\nabla_{i}X(x) \in W_{x}^{(0)}$ for all $i$,
by the definition of the derived flag. Since $\Lambda^{2}T^{*}M \subseteq W^{(0)}$,
$X(x) \in W_{x}^{(1)}$ if and only if 
\begin{eqnarray} \label{w}
K_{a;(i)} -L_{a (i)} \in ker \hspace{0.03in} \partial c
\end{eqnarray}
at $x$. Taking the covariant derivative of $c^{,s}K_{s}=0$ gives the equation
$ c^{,a}K_{a;b}=-c_{;ab}K^{a} $.
Substituting this into (\ref{w}) determines $W^{(1)}$ as the subset of
all $K+L \in W^{(0)}$ satisfying
\begin{eqnarray} \label{kl}
c_{;a b}K^{a}+L_{ab}c^{,a}=0 
\end{eqnarray}
By contracting (\ref{kl}) with $K$ and $\partial c$, it is evident that $W^{(1)}$ 
consists of the zero elements in $W$ along with the solutions of 
\begin{eqnarray} 
c_{;ab}K^{a}K^{b} + L_{ab}c^{,a}K^{b} = 0 \label{eqn1} \\
c_{;ab}K^{a}c^{,b} = 0 \label{eqn2}
\end{eqnarray}
where $0\neq K\in ker \hspace{0.03in} \partial c$ and $L\in \Lambda^{2}T^{*}M$.
Equation (\ref{eqn2}) has a solution  $0\neq K\in ker \hspace{0.03in} \partial c$ 
if and only if 
\begin{eqnarray} \label{cc}
c_{;ab}c^{,b} = f c_{a}
\end{eqnarray}
for some $f$. (\ref{cc}) may be written in terms of differential forms as
\begin{eqnarray} \label{diffform}
 dc\wedge D_{\partial c}dc =0 
\end{eqnarray}
where $D$ denotes covariant differentiation.
This leads to a necessary condition for the existence of a Killing field on
a Riemannian surface. 
\begin{theorem} If a regular Riemannian surface possesses
a Killing field then
\[  dc\wedge D_{\partial c}dc =0 \ \]
\end{theorem}
{\bf Proof:} \\
By Theorem \ref{thm:main}, if a regular Riemannian surface has a Killing field then
$\widetilde{W}$ has rank at least one. Since $\widetilde{W} \subseteq W^{(1)}$,
equation (\ref{eqn2}) must have a non-trivial
solution $K\in ker \hspace{0.03in} \partial c$ at each $x\in M$. The theorem now follows 
from the fact that (\ref{eqn2}) is equivalent to (\ref{diffform}). \\
{\bf q.e.d.}

\begin{corollary} Let $M$ be  a regular Riemannian surface with non-constant curvature. 
If $M$ possesses a Killing field then the integral 
curves of $\partial c$ are geodesic paths.
\end{corollary}
By a {\it geodesic path} we mean a curve that is a geodesic when appropriately
parameterized. \\
{\bf Proof:}\\
Equation (\ref{diffform}) is equivalent to $D_{\partial c}\partial c = f \partial c$, 
which implies that integral curves of the non-vanishing vector field $\partial c$ may be 
parametrized so as to give geodesics of $M$. \\
{\bf q.e.d.} 

An example would be the punctured paraboloid $z=x^{2}+y^{2}$;
$(x,y) \neq (0,0)$, with the induced metric from its embedding
into 3-dimensional Euclidean space. The integral curves of $\partial c$ are described
by the geodesic paths $\gamma_{\theta}(t) = (tcos\theta, tsin\theta, t^{2})$, up to 
reparametrization.

Now let us return to calculating $W^{(1)}$. If  $dc\wedge D_{\partial c}dc =0$ on
the surface then the non-zero elements of $W^{(1)}$ are the solutions to
(\ref{eqn1}) for which $0\neq K\in ker \hspace{0.03in} \partial c$. 
For any choice of non-trivial $K\in ker \hspace{0.03in} \partial c$,
(\ref{eqn1}) uniquely determines an element $L=L(K)\in \Lambda^{2}T^{*}M$. Therefore
$W^{(1)}$, in this case, is a rank one vector bundle over $M$. If, on the other hand,
$dc\wedge D_{\partial c}dc \neq 0$ on $M$ then $W^{(1)}$ is the zero bundle and there
do not exist any local Killing fields. As a consequence, ${\cal K}_{x}$ cannot be
2-dimensional; this may also be seen directly by considering the Lie bracket operation.
Henceforth we shall assume that $dc\wedge D_{\partial c}dc =0$. 
  
To find $W^{(2)}$, let $X=K+L$ be a local section of $W^{(1)}$. By definition,
$X(x) \in W_{x}^{(2)}$ if and only if $\nabla_{i}X(x) \in W^{(1)}_{x}$ for all $i$. 
Owing to (\ref{kl}), this is equivalent to
\begin{eqnarray} \label{eqn3}
c_{;a b}{{K'}^{a}}_{(i)}+{L'}_{ab(i)}c^{,a}=0 
\end{eqnarray}
where
\begin{eqnarray}
{K'}_{a(i)} & := & K_{a;(i)}-L_{a(i)}  \nonumber \\
{L'}_{ab(i)} & := & L_{ab;(i)} -{R_{ab(i)}}^{c}K_{c} \nonumber
\end{eqnarray}
(Note that from the description of $W^{(1)}$ contained in (\ref{w}) it follows that
${K'}_{a (i)} + {L'}_{ab(i)} \in W^{(0)}$.)
Taking the covariant derivative of (\ref{kl}) gives 
\begin{eqnarray} \label{covkl}
c_{;ab}{K^{a}}_{;(i)}+L_{ab;(i)}c^{,a}=-c_{;ab(i)}K^{a}-L_{ab}{c^{;a}}_{(i)}
\end{eqnarray}
Substituting (\ref{covkl}) into (\ref{eqn3}) defines $W^{(2)}$ as the subset of all
$K+L \in W^{(1)}$ such that
\begin{eqnarray} \label{eqn4}
c_{;abc}K^{a} + L_{ab}{c^{;a}}_{c}+L_{ac}{c^{;a}}_{b}+R_{abcd}c^{,a}K^{d} =0
\end{eqnarray} 
If this has only the trivial solution then $\widetilde{W}=W^{(2)}$ is the zero bundle. 
Otherwise, $\widetilde{W} = W^{(2)} = W^{(1)}$ has rank one. 

\begin{theorem} \label{thm:1diml} Let $M$ be a regular Riemannian surface.
Then $dim \hspace{0.03in} {\cal K}_{x}=1$ for all $x\in M$ if and only if \\
(i) \hspace*{0.06in} $dc\neq 0$, \\ 
(ii) \hspace*{0.01in} $dc\wedge D_{\partial c}dc =0$, \hspace{0.03in} and \\
(iii) equation (\ref{eqn4}) holds for all $K+L \in W^{(1)}$. 
\end{theorem}
{\bf Proof:}\\
Conditions (i)-(iii) are equivalent to $rank \hspace{0.02in} \widetilde{W} =1$. 
The result now follows from Theorem \ref{thm:main}. \\
{\bf q.e.d.} 

We summarize the discussion in this section with the following corollary.
\begin{corollary}
For regular Riemannian surfaces, ${\cal K}_{x}$ may be one of five possible 
Lie algebras. It is isomorphic to either ${\sf sl}_{2}\Re$ , ${\sf su}_{2}$ 
or the Lie algebra of $\Re^{2}\times_{sd} SO(2)$
when the Gaussian curvature is constant and positive, negative or zero, respectively. 
${\cal K}_{x}$ is
the 1-dimensional Lie algebra when the conditions of Theorem \ref{thm:1diml} are met. 
Otherwise, there do not exist any local Killing fields and ${\cal K}_{x}$ is trivial. 
\end{corollary}

\end{document}